\documentclass[aps,prl,reprint,groupedaddress,floatfix]{revtex4-1}
\usepackage{graphicx,amsmath,amssymb,amsfonts,mathtools,booktabs,array}
\usepackage[utf8]{inputenc}
\usepackage[T1]{fontenc}
\usepackage{lmodern}
\usepackage[colorlinks=true,linkcolor=blue,citecolor=blue,urlcolor=blue,breaklinks=true]{hyperref}
\usepackage{color}
\usepackage{graphicx}
\usepackage{microtype}
\DisableLigatures[f]{encoding=*,family=*}
\newcommand{\heading}[1]{\par\emph{#1}.---}
\newcommand{\E}{\mathbb E}
\newcommand{\Var}{\operatorname{Var}}
\newcommand{\Cov}{\operatorname{Cov}}
\newcommand{\ii}{\mathrm i}
\newcommand{\dd}{\,\mathrm d}

\newcommand{\post}{\mathrm{post}}
\newcommand{\tf}{\mathrm{tran}}

\newcommand{\Fano}{\mathcal F}
\newcommand{\rx}[1]{\xrightarrow{#1}}
\definecolor{boundblue}{RGB}{0,50,200}
\definecolor{boundred}{RGB}{177,50,48}
\definecolor{boundgreen}{RGB}{31,119,91}

\begin{document}
\title{Exact Stochastic Theory of Post-Transcriptional Gene Autoregulation Mediated by RNA-Binding Proteins}
\author{Chen Jia$^{1,2}$}
\email{Contact author: chenjia@hit.edu.cn}
\affiliation{$^1$ School of Mathematics, Harbin Institute of Technology, Harbin 150001, China}
\affiliation{$^2$ Beijing Research Institute, Harbin Institute of Technology, Beijing 100144, China}

\begin{abstract}
While analytical theories of transcriptional feedback loops have been well established, exact stochastic analyses of post-transcriptional feedback loops remain relatively scarce and are notoriously challenging. Here, we develop an exact theory of post-transcriptional gene autoregulation, in which reversible protein-mRNA binding modulates both translation and mRNA stability. We derive the state-dependent protein burst-size distribution, the steady-state protein number distribution in terms of confluent Heun functions, and an exact criterion for sub-Poissonian fluctuations. By comparing transcriptional and post-transcriptional networks, we reveal a reversal in their relative advantages in noise suppression and rhythm transmission: negative feedback favors post-transcriptional regulation, whereas positive feedback favors transcriptional regulation. Our results reveal how the regulatory pattern fundamentally shapes gene expression fluctuations and dynamic responses.
\end{abstract}

\maketitle

\heading{Introduction}
Feedback regulation is a fundamental organizing principle of living systems, enabling cells to maintain homeostasis, regulate molecular fluctuations, and coordinate dynamic responses to ever-changing environments \cite{alon2007network, brandman2008feedback}. A substantial body of quantitative theory has focused on transcriptional regulation, in which transcription factors bind to DNA and modulate promoter activity \cite{hornos2005self, ramos2007symmetry, visco2008exact, grima2012steady, kumar2014exact, jia2019single, jia2020small, wu2024solving}. However, post-transcriptional regulation is also widespread in nature, involving diverse regulatory proteins and noncoding RNAs that modulate mRNA processing, stability, and translation \cite{waters2009regulatory, cech2014noncoding, roundtree2017dynamic, hentze2018brave}. In recent years, several analytical studies have investigated how post-transcriptional regulation affects gene expression fluctuations \cite{jia2010post, kumar2018stochastic, fan2023effect, yu2026analytical, wong2026analyzing, aquino2012stochastic}. However, most of these studies have focused on regulation by external regulatory factors such as small noncoding RNAs and RNA-binding proteins, whereas endogenous feedback loops have received much less attention. This raises a fundamental question: how do post-transcriptional feedback loops shape gene expression fluctuations and dynamic responses?

Autoregulatory feedback loops are among the simplest gene regulatory networks \cite{becskei2000engineering}. Extensive theoretical studies have investigated transcriptional autoregulation, for which exact solutions have been obtained for several classical models \cite{hornos2005self, ramos2007symmetry, visco2008exact, grima2012steady, kumar2014exact, jia2019single, jia2020small, wu2024solving}. In contrast, post-transcriptional autoregulation mediated by RNA-binding proteins is also widespread in living cells \cite{muller2019auto}, whereby protein expressed from a gene binds to its own mRNA and regulates translation and/or mRNA stability. Compared with transcriptional feedback, post-transcriptional feedback involves more intricate coupling among protein-mRNA binding, translation, and mRNA degradation, posing substantial challenges to analytical treatment. Moreover, although previous work has suggested that translational negative feedback can suppress noise more effectively than transcriptional feedback \cite{swain2004efficient, bokes2017gene}, the relative advantages of these two regulatory mechanisms remain incompletely understood and have yet to be systematically compared.

In this study, we develop an analytical theory of post-transcriptional gene autoregulation. We derive exact protein burst-size and copy-number distributions, as well as an exact criterion for sub-Poissonian fluctuations. We also compare transcriptional and post-transcriptional feedback loops in terms of their abilities to suppress gene expression noise and transmit biological rhythms.

\begin{figure}[htb!]
\centering\includegraphics[width=\linewidth]{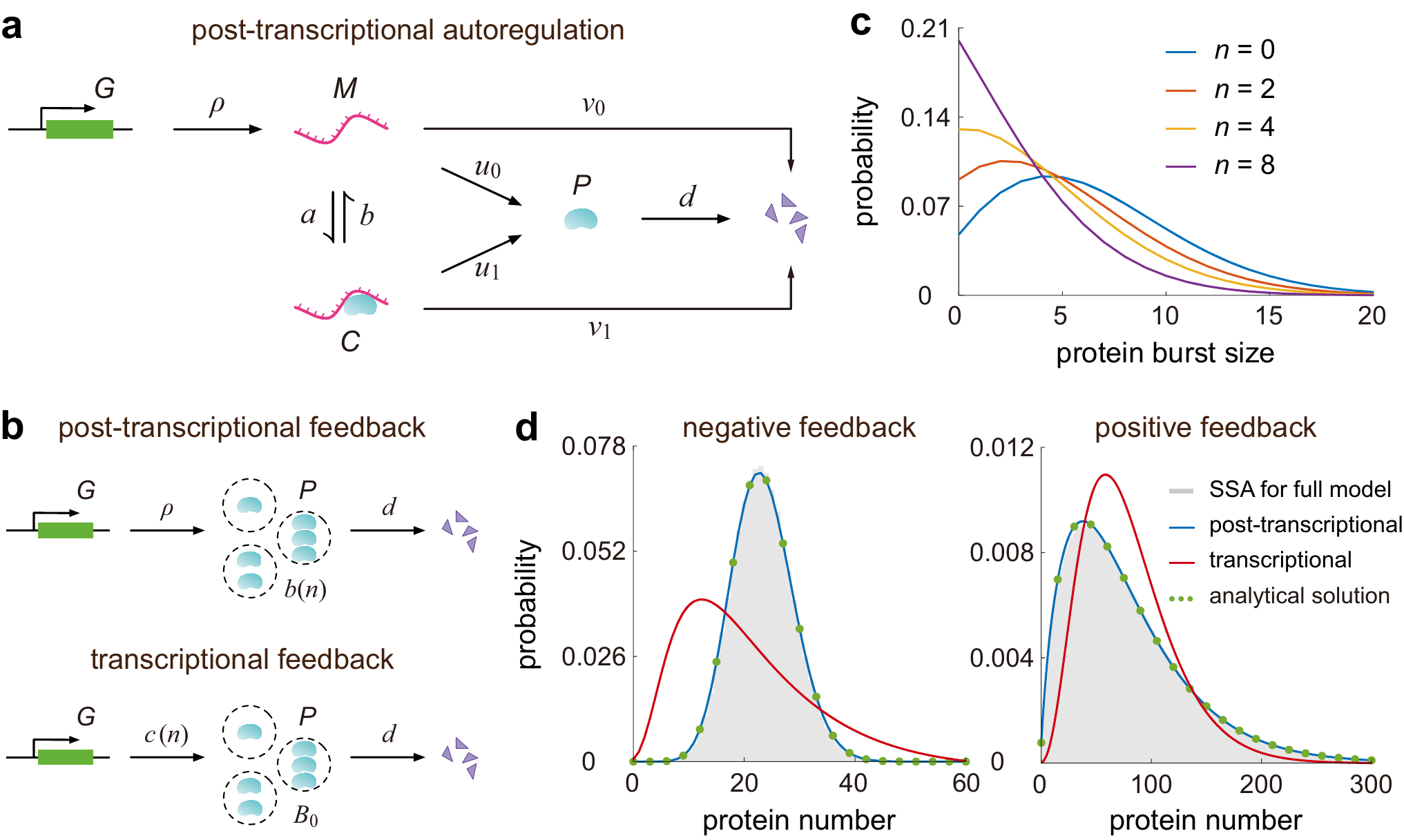}
\caption{\textbf{Stochastic dynamics of post-transcriptional autoregulation.}
\textbf{(a)} Schematic of the full model.
\textbf{(b)} Effective protein dynamics for post-transcriptional and transcriptional loops.
\textbf{(c)} Protein burst-size distribution in the post-transcriptional network.
\textbf{(d)} Comparison of the steady-state protein distributions in the two networks. The gray region represents the distribution of the full model in (a) obtained using the stochastic simulation algorithm (SSA), the blue and red curves represent the distributions of the effective models in (b) obtained using the SSA, and the green dots show the exact distribution given by Eq.~\eqref{eq:weights}. See Ref.~\cite{supp} for technical details of the figure.}\label{model}
\end{figure}

\heading{Model}
Consider a stochastic gene expression system with post-transcriptional autoregulation, in which protein can bind to its own mRNA and modulate both translation and mRNA degradation (Fig.~\ref{model}(a)). Let $G$ denote the gene of interest, and let $M$, $C$, and $P$ denote the free mRNA, mRNA-protein complex, and free protein, respectively. The reaction scheme is given by
\begin{equation}\label{eq:reactions}
\begin{gathered}
G\rx\rho G+M,\quad M+P\rx a C,\quad C\rx b M+P,\\
M\rx{u_0}M+P,\quad C\rx{u_1}C+P,\\
M\rx{v_0}\varnothing,\quad C\rx{v_1}P,\quad P\rx d\varnothing,
\end{gathered}
\end{equation}
where $\rho$ is the transcription rate, $d$ is the protein degradation rate, and $a$ and $b$ are the protein-mRNA binding and unbinding rates, respectively. Protein binding to mRNA can alter the translation rate from $u_0$ to $u_1$ and alter the mRNA degradation rate from $v_0$ to $v_1$. The reaction $C\to P$ accounts for the release of the bound protein upon degradation of the associated mRNA.

The microstate of the system can be described by the ordered triple $(m,c,n)$, where $m$, $c$, and $n$ denote the copy numbers of $M$, $C$, and $P$, respectively. Moreover, let $p_{m,c,n}$ denote the probability that the system is in microstate $(m,c,n)$. Then the stochastic gene expression dynamics is governed by the chemical master equation:
\begin{equation*}
\scalebox{0.91}{$\begin{split}
\dot p_{m,c,n}={}&
\rho p_{m-1,c,n}
+a(m+1)(n+1)p_{m+1,c-1,n+1}\\
&+b(c+1)p_{m-1,c+1,n-1}
+(u_0m+u_1c)p_{m,c,n-1}\\
&+v_0(m+1)p_{m+1,c,n}
+v_1(c+1)p_{m,c+1,n-1}\\
&+d(n+1)p_{m,c,n+1}-\big[\rho+m(an+u_0+v_0)\\
&+c(b+u_1+v_1)+dn\big]p_{m,c,n}.
\end{split}$}
\end{equation*}
At steady state, the mean synthesis and degradation fluxes of mRNA and protein must balance, yielding
\begin{equation}\label{eq:balances}
\rho=v_0\E m+v_1\E c,\qquad
d\E n=u_0\E m+u_1\E c,
\end{equation}
where $\E m$, $\E c$, and $\E n$ denote the steady-state means of $m$, $c$, and $n$, respectively. Combining these balance relations, we obtain
\begin{equation}\label{eq:meancomplex}
\E n=\frac{u_0\rho}{v_0d}
+\frac{v_1}{d}\left(\frac{u_1}{v_1}-\frac{u_0}{v_0}\right)\E c.
\end{equation}
Here, $u_0/v_0$ and $u_1/v_1$ represent the average numbers of protein molecules produced per free and bound mRNA, respectively. According to Eq.~\eqref{eq:meancomplex}, when $u_1/v_1>u_0/v_0$, protein binding to mRNA increases the steady-state mean protein number relative to the unregulated network. We refer to this as \emph{positive feedback}. When $u_1/v_1<u_0/v_0$, protein binding decreases the steady-state mean; this corresponds to \emph{negative feedback}. When $u_1/v_1=u_0/v_0$, protein binding has no effect on the steady-state mean, and the gene is effectively unregulated.

\heading{Effective protein bursts in the fast-mRNA limit}
In many cell types, including bacteria, yeast, and mammalian cells, mRNA typically has a much shorter lifetime than its protein counterpart \cite{shahrezaei2008analytical}. For example, in mouse fibroblasts, the median half-lives of mRNAs and proteins are approximately $9$ h and $46$ h, respectively \cite{schwanhausser2011global}. This timescale separation is even more pronounced in bacteria and yeast \cite{jia2021frequency}. In the fast-mRNA limit, protein synthesis occurs in stochastic bursts, allowing the full reaction system in Eq.~\eqref{eq:reactions} to be reduced to the following effective protein dynamics (Fig.~\ref{model}(b)):
\begin{equation}\label{eq:effective}
G\xrightarrow{\rho q_k(n)} G+kP,\quad k\geq 1,\qquad
P\xrightarrow{d}\varnothing.
\end{equation}
Here, $\rho$ is the transcription rate, which now serves as the protein burst frequency, and $q_k(n)$ denotes the protein burst-size distribution, which generally depends on the current protein number $n$. In Ref.~\cite{supp}, we derive the following exact expression for the burst-size distribution:
\begin{equation}\label{eq:qkernel}
q_k(n)=e_LT_n\cdots T_{n+k-1}(I-T_{n+k})e_R,
\quad k\geq0,
\end{equation}
where $e_L=(1,0)$, $e_R=(1,1)^T$, and
\begin{equation*}
T_j=\frac{1}{U_0U_1+bU_0+ajU_1}
\begin{bmatrix}
(U_1+b)u_0 & aju_1\\
bu_0 & (U_0+aj)u_1
\end{bmatrix},
\end{equation*}
are $2\times 2$ matrices with $U_0=u_0+v_0$ and $U_1=u_1+v_1$.

This highlights a fundamental distinction between post-transcriptional and transcriptional regulation. For transcriptional feedback, the protein burst size typically follows a geometric distribution independent of the current protein number $n$ \cite{kumar2014exact, jia2020small}. In contrast, post-transcriptional feedback generally gives rise to a non-geometric burst-size distribution that depends explicitly on $n$. Fig.~\ref{model}(c) shows how the burst-size distribution varies with $n$. Clearly, the distribution is approximately geometric for large $n$, whereas it may deviate substantially from the geometric distribution for small $n$.

To understand the origin of this effective description, note that each newly synthesized mRNA initiates a brief episode of translation and reversible protein binding, which terminates upon mRNA degradation. On the protein timescale, this episode can be treated as instantaneous. Since protein binding and unbinding conserve the total protein number, and degradation of a bound mRNA releases its associated protein, these processes do not contribute to the net change in protein number. Consequently, the net protein increment during each episode, i.e. the protein burst size, is simply the number of translation events occurring over the mRNA lifetime, whose distribution is given by $q_k(n)$ \cite{supp}.

In particular, when $u_0/v_0=u_1/v_1$, the gene is effectively unregulated. In this case, it is straightforward to verify that $T_je_R=pe_R$ and $(I-T_j)e_R=(1-p)e_R$, where $p=u_0/(u_0+v_0)=u_1/(u_1+v_1)$. Substituting these identities into Eq.~\eqref{eq:qkernel} yields a geometric burst-size distribution,
$q_k=p^k(1-p)$, which is independent of the current protein number $n$. Hence, the steady-state protein number follows the well-known negative binomial distribution $p_n^{\mathrm{ss}}=\frac{(r)_n}{n!}p^n(1-p)^r$ \cite{paulsson2000random}, where $r=\rho/d$ denotes the mean number of bursts per protein lifetime.

\heading{Exact steady-state protein distribution}
We now derive the steady-state protein number distribution for the effective dynamics in Eq.~\eqref{eq:effective}. Let $p_n^{\mathrm{ss}}$ denote the steady-state probability of having $n$ protein molecules, and define the generating function $G(z)=\sum_{n=0}^{\infty}p_n^{\mathrm{ss}}z^n$. Although the burst-size distribution is expressed in terms of matrix products, its underlying two-state structure allows us to derive a closed second-order differential equation for the generating function $G(z)$ (see End Matter):
\begin{align}\label{eq:pgfode}
&az(u_1+v_1-u_1z)G''(z)+\bigl\{U_0(U_1+b) \nonumber\\
&-[au_1(1+r)+bu_0+u_0U_1+u_1U_0]z+u_0u_1z^2\bigr\}G'(z) \nonumber\\
&+ru_0(u_1z-b-U_1)G(z)=0.
\end{align}
The physical solution must be analytic at $z=0$ and satisfy the normalization condition $G(1)=1$.

For $u_1>0$, Eq.~\eqref{eq:pgfode} has two finite regular singular points, at $z=0$ and $z=1+v_1/u_1>1$, and can therefore be transformed into a confluent Heun equation. To see this, we introduce $\zeta=u_1/(u_1+v_1)$ and perform the following change of variables:
\begin{equation}
x=\zeta z,\qquad H(x)=G(z)=G(x/\zeta).
\end{equation}
The resulting equation for $H(x)$ takes the standard confluent Heun form:
\begin{equation}\label{eq:heuncanonical}
H''+\left(\frac{\gamma_H}{x}+\frac{\delta_H}{x-1}+\epsilon_H\right)H'
+\frac{A_Hx-q_H}{x(x-1)}H=0,
\end{equation}
with parameters
\begin{equation*}\label{eq:heunparameters}
\begin{gathered}
A_H=-\frac{ru_0U_1}{au_1},\quad
q_H=-\frac{ru_0(U_1+b)}{au_1},\quad
\epsilon_H=-\frac{u_0U_1}{au_1},\\
\gamma_H=\frac{U_0(U_1+b)}{aU_1},\quad
\delta_H=1+r+\frac{b}{a}
\left(\frac{u_0}{u_1}-\frac{U_0}{U_1}\right).
\end{gathered}
\end{equation*}
The solution analytic at $x=0$ is then given by the confluent Heun function:
\begin{equation}
H(x)=K\operatorname{HeunC}(q_H,A_H,\gamma_H,\delta_H,\epsilon_H;x),
\end{equation}
where $K$ is a normalization constant. Imposing $G(1)=1$, we obtain the steady-state generating function:
\begin{equation}
G(z)=H(\zeta z)=
\frac{\operatorname{HeunC}(q_H,A_H,\gamma_H,\delta_H,\epsilon_H;\zeta z)}
{\operatorname{HeunC}(q_H,A_H,\gamma_H,\delta_H,\epsilon_H;\zeta)}.
\end{equation}
The steady-state protein number distribution can then be recovered from the derivatives of $G(z)$ at the origin:
\begin{equation}\label{eq:derivatives}
p_n^{\mathrm{ss}}=\frac{1}{n!}
\left.\frac{\dd^nG(z)}{\dd z^n}\right|_{z=0}.
\end{equation}

Alternatively, the steady-state protein distribution admits the following representation in terms of a sequence of recursively defined weights (see End Matter):
\begin{equation}\label{eq:weights}
p_n^{\mathrm{ss}}=\frac{w_n}{\sum_{j=0}^{\infty}w_j},
\end{equation}
where the weights $w_n$, $n\geq0$, satisfy the three-term recurrence relation
\begin{equation}\label{eq:recurrence}
\begin{aligned}
&(n+1)[U_0(U_1+b)+anU_1]w_{n+1}\\
&=\bigl\{[anu_1+(U_1+b)u_0](n+r)+nu_1U_0\bigr\}w_n\\
&\quad-u_0u_1(n+r-1)w_{n-1},
\qquad n\geq0,
\end{aligned}
\end{equation}
with $w_{-1}=0$ and $w_0=1$. Eqs.~\eqref{eq:derivatives} and \eqref{eq:weights} provide two equivalent characterizations of the same protein distribution. As expected, the analytical solution coincides perfectly with numerical simulations (Fig.~\ref{model}(d)).

\heading{An exact sub-Poisson criterion for complete translation repression}
The protein distribution takes a simpler form when protein binding to mRNA completely represses translation, i.e. $u_1=0$ \cite{swain2004efficient}. In this case, the burst-size distribution in Eq.~\eqref{eq:qkernel} reduces to
\begin{equation}\label{eq:specialburst}
q_k(n)=\frac{\beta^k(n+\alpha+k-\beta)}{(n+\alpha)_{k+1}},
\end{equation}
and Eq.~\eqref{eq:pgfode} also reduces to the confluent hypergeometric differential equation
\begin{equation}\label{eq:kummerode}
zG''(z)+(\alpha-\beta z)G'(z)-r\beta G(z)=0,
\end{equation}
where $(x)_k=x(x+1)\cdots(x+k-1)$ denotes the rising factorial, and
\begin{equation}\label{eq:kummerparameters}
r=\frac{\rho}{d},\quad
\eta=\frac{av_1}{b+v_1},\quad
\alpha=\frac{u_0+v_0}{\eta},\quad
\beta=\frac{u_0}{\eta}.
\end{equation}
The steady-state solution of Eq.~\eqref{eq:kummerode} is given by
\begin{equation}\label{eq:kummer}
\begin{gathered}
G(z)=\frac{M(r;\alpha;\beta z)}{M(r;\alpha;\beta)},\quad
p_n^{\mathrm{ss}}=\frac{(r)_n\beta^n}{(\alpha)_n n!\,M(r;\alpha;\beta)},
\end{gathered}
\end{equation}
where $M(r;\alpha;x)={}_1F_1(r;\alpha;x)$ denotes Kummer's confluent hypergeometric function.

In single-cell experiments, protein number fluctuations around the mean are often characterized by the Fano factor $\Fano=\Var(n)/\E n$. Sub-Poissonian fluctuations, characterized by a Fano factor smaller than $1$, are often regarded as a hallmark of effective noise suppression. For complete translation repression ($u_1=0$), we obtain the \emph{exact sub-Poissonian criterion} (see End Matter):
\begin{equation}\label{eq:subpoisson}
\Fano<1\ \Longleftrightarrow\ r>\alpha.
\end{equation}
In terms of the original biochemical parameters, the condition $r>\alpha$ becomes
\begin{equation}\label{eq:biochemicalthreshold}
\frac{\rho}{d}>\frac{(u_0+v_0)(b+v_1)}{av_1}.
\end{equation}
Hence, stronger protein-mRNA binding and more frequent mRNA production facilitate noise suppression, whereas faster protein-mRNA dissociation, translation, and protein degradation weaken this effect.

\heading{Comparison with transcriptional autoregulation}
A natural and important question is: what advantages does post-transcriptional regulation offer over transcriptional regulation? To address this question, we next compare the post-transcriptional feedback model in Eq.~\eqref{eq:effective} with a transcriptional feedback loop described by the following effective reactions (Fig.~\ref{model}(b)) \cite{friedman2006linking, mackey2013dynamic, bokes2013transcriptional, jia2017stochastic, bokes2022exact}:
\begin{equation}\label{eq:transcriptional}
G\xrightarrow{c(n)q_k} G+kP,\quad k\geq 1,\qquad
P\xrightarrow{d}\varnothing.
\end{equation}
Here, $c(n)$ is the burst frequency, which depends on the protein number $n$, and the burst size is assumed to follow a geometric distribution $q_k$ with mean $B_0=u_0/v_0$. Clearly, the protein number regulates the burst frequency in the transcriptional network, whereas it regulates the burst size in the post-transcriptional network.

To ensure a fair comparison, we take $c(n) = \rho b(n)/B_0$, where $b(n)=\sum_{k=0}^\infty kq_k(n)$ is the mean burst size in the post-transcriptional feedback loop \cite{bokes2017gene}. With this choice, both networks have the same conditional mean protein synthesis rate, $a(n)=\rho b(n)=B_0c(n)$, and therefore share the same mean-field (deterministic) dynamics:
\begin{equation}\label{eq:driftmatch}
\dot n=a(n)-dn.
\end{equation}

We next compare the noise-suppression capabilities of the two networks. To this end, we employ the linear noise approximation (LNA) \cite{grima2015linear} around the common stable fixed point $n_*$ of Eq.~\eqref{eq:driftmatch}, satisfying $a(n_*)=dn_*$. Under the LNA, the protein number in either network can be written as $n=n_*+\xi_X$, where the fluctuation $\xi_X$ satisfies the linear Langevin equation \cite{supp}
\begin{equation}
\mathrm d\xi_X=-\kappa\xi_X\mathrm dt+\sqrt{D_X}\mathrm d\mathcal W_X,
\end{equation}
where the subscript $X\in\{\post,\tf\}$ indicates whether the network is post-transcriptional or transcriptional, $\mathcal W_X$ is a standard Wiener process, and $\kappa=d-\rho b'(n_*)>0$ is the relaxation rate toward the steady state. The diffusion coefficients are given by
\begin{equation}\label{eq:diffusions}
D_\post=dn_*+\rho s(n_*),\quad
D_\tf=2dn_*(1+B_0),
\end{equation}
where $s(n)=\sum_{k=0}^\infty k^2q_k(n)$ is the second moment of the burst-size distribution in the post-transcriptional network.

Since the LNA approximates the discrete protein dynamics by a Langevin equation, the variance and Fano factor of protein fluctuations can be readily obtained as
\begin{equation}
\Var_X(n)\approx\frac{D_X}{2\kappa},\quad
\Fano_X\approx\frac{D_X}{2\kappa n_*}.
\end{equation}
Consequently, the difference between the Fano factors of the two networks is given by
\begin{equation}\label{eq:noisedifference}
\Fano_\post-\Fano_\tf
\approx\frac{D_\post-D_\tf}{2\kappa n_*}.
\end{equation}
In \cite{supp}, we establish the exact relation
\begin{equation}\label{eq:keyrelation}
D_\post-D_\tf=2\rho\theta W(n_*),
\end{equation}
where $\theta=u_1/v_1-u_0/v_0$ and $W(n_*)>0$ measures the average number of protein molecules produced per burst in the post-transcriptional network, counting only those bursts that terminate with degradation of the protein-bound mRNA. From Eqs.~\eqref{eq:noisedifference} and \eqref{eq:keyrelation}, the difference in the noise-suppression capabilities of the two networks is entirely determined by the sign of $\theta$. For negative feedback ($\theta<0$), we have $D_\post<D_\tf$, indicating that post-transcriptional regulation suppresses gene expression noise more effectively than transcriptional regulation. For positive feedback ($\theta>0$), however, the opposite holds, and transcriptional regulation provides stronger noise suppression (Fig.~\ref{model}(d)).

We emphasize that the noise-suppression advantage of post-transcriptional regulation has been previously identified in Ref.~\cite{swain2004efficient} for negative autoregulation with complete translation repression ($u_1=0$). Here, we not only extend this result to the more general case of $u_1>0$, but also reveal a reversal of this advantage under positive autoregulation: in terms of noise suppression, negative feedback favors post-transcriptional regulation, whereas positive feedback favors transcriptional regulation.

\heading{Transmission of upstream biological rhythms}
Gene expression in living cells is frequently driven by rhythmic upstream signals, such as circadian clocks, cell-cycle progression, and pulsatile regulators including NF-$\kappa$B and p53 \cite{nicolas2018modulation, skinner2016single, lahav2004dynamics, nelson2004oscillations}. However, rhythmic upstream signals do not necessarily give rise to rhythmic downstream gene expression responses. We next compare the abilities of the two networks to transmit upstream biological rhythms.

To this end, we model the upstream rhythmic signal as an arbitrary stationary stochastic process $S_t$ with zero mean and finite second moments. Moreover, we assume that the burst frequencies of the two networks are modulated by the upstream signal according to $\rho(1+\epsilon S_t)$ and $c(n)(1+\epsilon S_t)$, respectively, where $\epsilon>0$ is the modulation strength. For simplicity, we assume weak modulation, i.e. $\epsilon$ is small. Under the weak-modulation assumption and the LNA, the protein dynamics in either network can be represented by $n=n_*+\xi_X$, where $\xi_X$ satisfies the linear non-autonomous Langevin equation
\begin{equation}\label{eq:drivenlna}
\mathrm d\xi_X=-\kappa\xi_X\mathrm dt
+\epsilon dn_*S_t\mathrm dt+\sqrt{D_X}\mathrm d\mathcal W_X.
\end{equation}

Stochastic rhythms are commonly characterized by autocovariance functions and power spectra \cite{jia2021frequency, gupta2022frequency}. For the upstream rhythmic signal $S_t$ and downstream protein level $n_t$, their steady-state autocovariance functions are defined as $C_S(t)=\Cov(S_0,S_t)$ and $C_X(t)=\Cov(n_0,n_t)$, where $\Cov(\cdot,\cdot)$ denotes covariance and $X\in\{\post,\tf\}$ indicates the regulation type. The corresponding power spectra are given by the Fourier transforms
\begin{equation*}
G_S(\omega)=\int_{\mathbb R}C_S(t)e^{-\ii\omega t}\dd t,\quad
G_X(\omega)=\int_{\mathbb R}C_X(t)e^{-\ii\omega t}\dd t,
\end{equation*}
where $\omega\geq 0$ denotes the angular frequency. In general, a nonmonotonic spectrum with a peak at a nonzero frequency provides a characteristic signature of stochastic rhythms, with the peak location identifying the dominant frequency. For the LNA in Eq.~\eqref{eq:drivenlna}, the power spectrum of downstream gene expression is given by \cite{supp}
\begin{equation}\label{eq:psd}
G_X(\omega)=\frac{\epsilon^2d^2n_*^2G_S(\omega)}{\kappa^2+\omega^2}+\frac{D_X}{\kappa^2+\omega^2},
\end{equation}
where the first term on the right-hand side describes the contribution of the upstream drive, while the second term represents the contribution of intrinsic gene expression noise, which acts as a low-frequency background.

\begin{figure}[htb!]
\centering\includegraphics[width=\linewidth]{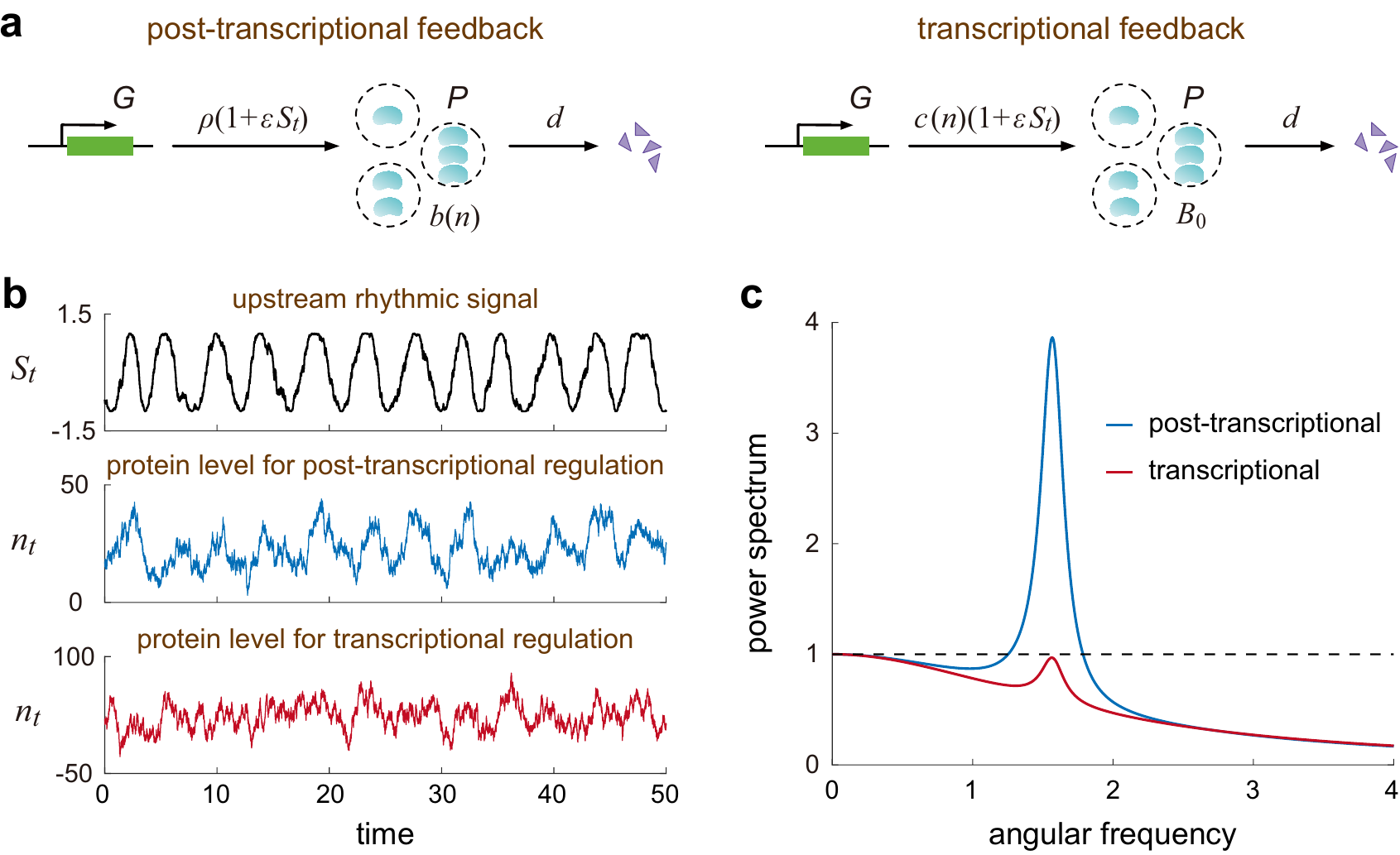}
\caption{\textbf{Rhythm transmission in post-transcriptional and transcriptional networks.}
\textbf{(a)} Schematics of the two networks, in which the protein burst frequency is modulated by an upstream rhythmic signal.
\textbf{(b)} Representative trajectories of the upstream and downstream signals under negative feedback. The upstream signal is modeled as a phase-diffusion oscillator $S_t=\cos\phi_t$, with $\mathrm d\phi_t=\omega_u\,\mathrm dt+\sqrt{2D_u}\,\mathrm dW_t$ and a uniformly distributed initial phase. Here, $\omega_u$ determines the mean oscillation frequency, while $D_u$ controls the variability in oscillation timing.
\textbf{(c)} Power spectra of protein number fluctuations in the two networks. See Ref.~\cite{supp} for technical details of the figure.}\label{transmission}
\end{figure}

Two conclusions can be drawn from Eq.~\eqref{eq:psd}. First, the difference in rhythm transmission between the two networks arises entirely from their intrinsic noise backgrounds (Fig.~\ref{transmission}(b),(c)). For negative feedback, we have $D_\post<D_\tf$, indicating that post-transcriptional regulation produces a lower intrinsic noise background and hence a higher signal-to-noise ratio, thereby facilitating rhythm transmission. For positive feedback, however, we have $D_\post>D_\tf$ and the advantage is reversed: transcriptional regulation provides a higher signal-to-noise ratio and thus facilitates rhythm transmission. Thus, the feedback mechanism that more effectively suppresses noise also more effectively preserves the rhythmic information encoded in the upstream signal.

Second, recall that $\kappa=d-\rho b'(n_*)>0$ represents the relaxation rate toward the steady state. Previous studies have shown that negative autoregulation speeds up relaxation, whereas positive autoregulation slows it down \cite{rosenfeld2002negative, jia2018relaxation}. Compared with positive feedback, negative feedback gives rise to a larger $\kappa$, which shortens the response time and reduces the phase lag between upstream and downstream signals, thereby enabling downstream gene expression to more faithfully track upstream rhythmic dynamics (Supplemental Fig.~S1).

\heading{Conclusion}
Here, we have developed an analytical theory of post-transcriptional autoregulation based on mesoscopic biochemical reaction dynamics. We derived the exact protein burst-size distribution, characterized the steady-state protein number distribution in terms of confluent Heun functions, and established an exact criterion for sub-Poissonian fluctuations. By comparing post-transcriptional and transcriptional feedback loops, we revealed a striking reversal of their noise-suppression and rhythm-transmission advantages: negative feedback favors post-transcriptional regulation, whereas positive feedback favors transcriptional regulation.

\heading{Acknowledgments}
The author acknowledges support from the National Natural Science Foundation of China under Grant Nos. 12526215, T2541031, and 12271020.

\footnotesize\setlength{\bibsep}{5pt}
\bibliography{references}
\bibliographystyle{apsrev4-2}

\normalsize
\section*{End Matter}

\heading{Derivation of the generating-function equation}
We now derive the differential equation for the steady-state generating function $G(z)=\sum_{n=0}^{\infty}p_n^{\mathrm{ss}}z^n$ using the effective protein dynamics in Eq.~\eqref{eq:effective} and the exact burst-size distribution in Eq.~\eqref{eq:qkernel}. Let $J_n$ denote the number of proteins produced in a single burst initiated with $n$ protein molecules. From Eq.~\eqref{eq:qkernel}, the survival function of $J_n$ is given by
\begin{equation}\label{eq:pgf-burst-tail}
Q_k(n):=\Pr(J_n\geq k)
=e_L T_n\cdots T_{n+k-1}e_R.
\end{equation}
Here the product of the transition matrices accounts for the first $k$ translation events, while multiplication by $e_R$ sums over the two possible mRNA states (free and bound) following the $k$th transition event.

Our derivation begins with a steady-state flux balance equation across the boundary separating states with at most $n$ protein molecules from those with at least $n+1$ protein molecules. Since protein degradation removes one protein molecule at a time, the downward flux across this boundary is $d(n+1)p_{n+1}^{\mathrm{ss}}$. In contrast, an upward crossing occurs whenever a burst initiated in state $i\leq n$ produces at least $n+1-i$ proteins. The corresponding upward flux is $\rho p_i^{\mathrm{ss}}Q_{n+1-i}(i)$. Equating the total upward and downward fluxes yields
\begin{equation}\label{eq:pgf-boundary-balance}
d(n+1)p^{\mathrm{ss}}_{n+1}
=\rho\sum_{i=0}^{n}p^{\mathrm{ss}}_iQ_{n+1-i}(i).
\end{equation}

To exploit the matrix-product structure of the burst-size distribution, we introduce the row vector
\begin{equation}\label{eq:pgf-crossing-vector}
h_n:=\sum_{i=0}^{n}p^{\mathrm{ss}}_i e_L T_i\cdots T_n,
\end{equation}
with $h_{-1}=(0,0)$. Its two components represent the contributions to the upward probability flux across the boundary from $n$ to $n+1$ protein molecules, distinguished by whether the mRNA is free or bound immediately after the upward crossing. Multiplying their sum by the burst frequency $\rho$ gives the total upward probability flux. Summing these contributions and using Eq.~\eqref{eq:pgf-boundary-balance}, we obtain
\begin{equation}\label{eq:pgf-crossing-current}
h_ne_R=\frac{n+1}{r}p^{\mathrm{ss}}_{n+1},
\end{equation}
where $r=\rho/d$. Moreover, separating the term $i=n$ in the definition of $h_n$ gives the recursive identity
\begin{equation}\label{eq:pgf-vector-recursion}
h_n=(p^{\mathrm{ss}}_ne_L+h_{n-1})T_n.
\end{equation}
This recursion is central to the derivation: it incorporates all preceding matrix products into $h_{n-1}$, leaving only the transition matrix $T_n$ associated with the next translation event.

We first consider the case of $u_1>0$, for which the transition matrix $T_n$ is invertible. It is easy to check that the inverse of $T_n$ is given by
\begin{equation}\label{eq:pgf-inverse-matrix}
T_n^{-1}
=\begin{bmatrix}
(U_0+an)/u_0 & -an/u_0\\
-b/u_1 & (U_1+b)/u_1
\end{bmatrix},
\end{equation}
where $U_0=u_0+v_0$ and $U_1=u_1+v_1$. For convenience, we write $h_n=(h_n^F,h_n^C)$. Then we can express the recursion relation in Eq.~\eqref{eq:pgf-vector-recursion} as $h_nT_n^{-1}=p_n^{\mathrm{ss}}e_L+h_{n-1}$, or equivalently,
\begin{equation}\label{eq:pgf-component-recursion}
\begin{aligned}
\frac{U_0+an}{u_0}h_n^F-\frac{b}{u_1}h_n^C
&=p^{\mathrm{ss}}_n+h_{n-1}^F,\\
-\frac{an}{u_0}h_n^F+\frac{U_1+b}{u_1}h_n^C
&=h_{n-1}^C.
\end{aligned}
\end{equation}

We now introduce the generating functions for the two components of $h_n$:
\begin{equation}\label{eq:pgf-component-functions}
A(z)=\sum_{n=0}^{\infty}h_n^Fz^n,
\qquad
B(z)=\sum_{n=0}^{\infty}h_n^Cz^n.
\end{equation}
Multiplying Eq.~\eqref{eq:pgf-component-recursion} by $z^n$ and summing over $n\geq0$, we obtain the coupled equations
\begin{equation}\label{eq:pgf-component-system}
\begin{aligned}
azA'(z)+(U_0-u_0z)A(z)
-\frac{bu_0}{u_1}B(z)&=u_0G(z),\\
-\frac{au_1}{u_0}zA'(z)
+(U_1+b-u_1z)B(z)&=0.
\end{aligned}
\end{equation}
The flux-balance relation in Eq.~\eqref{eq:pgf-crossing-current} provides an additional identity:
\begin{equation}\label{eq:pgf-component-sum}
A(z)+B(z)=\frac{G'(z)}{r}.
\end{equation}
These equations allow us to eliminate the auxiliary generating functions $A(z)$ and $B(z)$ and obtain a closed equation for $G(z)$.

To this end, multiplying the second equation in Eq.~\eqref{eq:pgf-component-system} by $u_0/u_1$ and adding it to the first eliminates $A'(z)$, yielding
\begin{equation}\label{eq:pgf-component-algebraic}
(U_0-u_0z)A(z)
+\frac{u_0}{u_1}(U_1-u_1z)B(z)
=u_0G(z).
\end{equation}
Let $H_0=u_0v_1-u_1v_0=u_0U_1-u_1U_0$. Here we focus on the regulated case of $H_0\neq0$, since $H_0=0$ corresponds to the unregulated model which has already been discussed in the main text. Combining Eqs.~\eqref{eq:pgf-component-sum} and \eqref{eq:pgf-component-algebraic}, we obtain
\begin{equation}\label{eq:pgf-component-elimination}
\begin{aligned}
A(z)
&=\frac{u_0}{rH_0}
\left[(U_1-u_1z)G'(z)-ru_1G(z)\right],\\
B(z)
&=\frac{u_1}{rH_0}
\left[ru_0G(z)-(U_0-u_0z)G'(z)\right].
\end{aligned}
\end{equation}
Differentiating the expression for $A(z)$, we find
\begin{equation}\label{eq:pgf-component-derivative}
A'(z)=\frac{u_0}{rH_0}
\left[(U_1-u_1z)G''(z)-u_1(1+r)G'(z)\right].
\end{equation}
Substituting these expressions into the second equation of Eq.~\eqref{eq:pgf-component-system} and simplifying, we arrive at
\begin{align}
0={}&az(U_1-u_1z)G''(z) \nonumber\\
&+\left[(U_0-u_0z)(U_1+b-u_1z)
-au_1(1+r)z\right]G'(z) \nonumber\\
&-ru_0(U_1+b-u_1z)G(z). \label{eq:pgf-factorized-derivation}
\end{align}
Finally, expanding the coefficients and using $U_1=u_1+v_1$, we obtain
\begin{align}
&az(u_1+v_1-u_1z)G''(z)
+\bigl\{U_0(U_1+b)\nonumber\\
&\quad-[au_1(1+r)+bu_0+u_0U_1+u_1U_0]z
+u_0u_1z^2\bigr\}G'(z)\nonumber\\
&\quad+ru_0(u_1z-b-U_1)G(z)=0,
\end{align}
which recovers Eq.~\eqref{eq:pgfode} in the main text.

\heading{Derivation of the three-term recurrence relation}
We now derive the recurrence relation for the steady-state protein distribution from the generating-function equation derived above. Recall that the first two derivatives of the generating function $G(z)$ are given by
\begin{equation}
\begin{aligned}
G'(z)&=\sum_{n=0}^{\infty}(n+1)p_{n+1}^{\mathrm{ss}}z^n,\\
G''(z)&=\sum_{n=0}^{\infty}(n+1)(n+2)p_{n+2}^{\mathrm{ss}}z^n.
\end{aligned}
\end{equation}
Substituting these expansions into the differential equation in Eq.~\eqref{eq:pgfode} and collecting the coefficients of $z^n$, we obtain
\begin{align}
&\left[an(n+1)U_1
+(n+1)U_0(U_1+b)\right]p_{n+1}^{\mathrm{ss}}
\nonumber\\
&-\big\{au_1n(n-1)
+n[au_1(1+r)+bu_0+u_0U_1+u_1U_0]
\nonumber\\
&+ru_0(U_1+b)\big\}p_n^{\mathrm{ss}}
+u_0u_1(n+r-1)p_{n-1}^{\mathrm{ss}}=0,
\end{align}
for $n\geq0$, where we set $p_{-1}^{\mathrm{ss}}=0$. Rearranging terms, we obtain the following three-term recurrence relation:
\begin{equation}\label{eq:originalrec}
\begin{aligned}
&(n+1)[U_0(U_1+b)+anU_1]p_{n+1}^{\mathrm{ss}}\\
&=\bigl\{[anu_1+(U_1+b)u_0](n+r)
+nu_1U_0\bigr\}p_n^{\mathrm{ss}}\\
&\quad-u_0u_1(n+r-1)p_{n-1}^{\mathrm{ss}},
\qquad n\geq0.
\end{aligned}
\end{equation}
Since this recurrence relation is homogeneous, the steady-state probabilities are determined up to a common multiplicative constant. We thus introduce the unnormalized weights
\begin{equation}
w_n=p_n^{\mathrm{ss}}/p_0^{\mathrm{ss}},
\end{equation}
with $w_{-1}=0$ and $w_0=1$. From Eq.~\eqref{eq:originalrec}, the unnormalized weights $w_n$ satisfy Eq.~\eqref{eq:recurrence} in the main text. Imposing the normalization condition $\sum_{n=0}^{\infty}p_n^{\mathrm{ss}}=1$ then yields Eq.~\eqref{eq:weights} in the main text. This provides an alternative characterization of the steady-state protein distribution through a three-term recurrence relation, without requiring the evaluation of confluent Heun functions.

\heading{Derivation of the exact sub-Poissonian criterion}
We now establish the exact criterion for sub-Poissonian protein fluctuations in the case of complete translation repression ($u_1=0$). Recall that the steady-state protein distribution is given by
\begin{equation}
p_n^{\mathrm{ss}}=
\frac{(r)_n\beta^n}{(\alpha)_n n!\,M(r;\alpha;\beta)}.
\end{equation}
Taking the ratio of two consecutive probabilities yields
\begin{equation}\label{eq:subpoisson-ratio}
(n+1)p_{n+1}^{\mathrm{ss}}
=\lambda(n)p_n^{\mathrm{ss}},
\qquad
\lambda(n)=\frac{\beta(n+r)}{n+\alpha}.
\end{equation}
Multiplying this identity by $1$ and $n$, respectively, and summing over $n\geq0$, we obtain
\begin{equation}\label{eq:subpoisson-moments}
\begin{aligned}
\E n&=\E[\lambda(n)],\\
\E[n(n-1)]&=\E[n\lambda(n)],
\end{aligned}
\end{equation}
where all expectations are taken with respect to the steady-state protein distribution. Consequently,
\begin{equation}\label{eq:subpoisson-covariance}
\begin{aligned}
\Var(n)-\E n
&=\E[n(n-1)]-(\E n)^2\\
&=\E[n\lambda(n)]-\E n\,\E[\lambda(n)]\\
&=\Cov(n,\lambda(n)).
\end{aligned}
\end{equation}
The key observation is that
\begin{equation}
\lambda'(n)=\frac{\beta(\alpha-r)}{(n+\alpha)^2}.
\end{equation}
Thus, $\lambda(n)$ is strictly decreasing when $r>\alpha$, strictly increasing when $r<\alpha$, and constant when $r=\alpha$. Since the covariance between a nondegenerate random variable and a strictly decreasing (increasing) function of itself is strictly negative (positive), Eq.~\eqref{eq:subpoisson-covariance} implies
\begin{equation}
\begin{aligned}
\Var(n)<\E n &\iff r>\alpha,\\
\Var(n)=\E n &\iff r=\alpha,\\
\Var(n)>\E n &\iff r<\alpha.
\end{aligned}
\end{equation}
Recalling that $\Fano=\Var(n)/\E n$, we thus obtain the exact sub-Poissonian criterion
\begin{equation}
\Fano<1\ \Longleftrightarrow\ r>\alpha,
\end{equation}
which establishes Eq.~\eqref{eq:subpoisson} in the main text.

\end{document}